\def\etal{{\it et al.\thinspace}}
\def\mearth{{\rm\,M_\oplus}}
\def\msun{{\rm\,M_\odot}}

\documentclass[useAMS,preprint]{mn2e}

\usepackage[dvips]{graphics}
\usepackage{epsfig}
\usepackage{deluxetable} 
\usepackage{amsmath}

\begin{document}

%\shorttitle{Formation of hot Earths}
%\shortauthors{Raymond, Barnes \& Mandell}

%\title{Observable Consequences of Planet Formation Models in Systems
%with Close-in Terrestrial Planets}

%\author{Sean N. Raymond\altaffilmark{1,\dagger,\star}, Rory
%Barnes\altaffilmark{2}, and Avi M. Mandell \altaffilmark{3,\dagger,\star}}
%\altaffiltext{1}{Center for Astrophysics and Space Astronomy, University of
%Colorado, Boulder, CO 80309-0389; raymond@lasp.colorado.edu}
%\altaffiltext{2}{Lunar and Planetary Laboratory, University of Arizona, Tucson, AZ 85721}
%\altaffiltext{3}{NASA Goddard Space Flight Center, Greenbelt, MD 20771}
%\altaffiltext{$\dagger$}{NASA Postdoctoral Program Fellow}
%\altaffiltext{$\star$}{Member of NASA Astrobiology Institute}

\title[Observable Consequences of Planet Formation Models]{Observable Consequences of Planet Formation Models in Systems
with Close-in Terrestrial Planets}
\author[Raymond, Barnes \& Mandell]{Sean
N. Raymond$^{1\dagger\star}$, Rory
Barnes$^2$, Avi M. Mandell$^{3\dagger\star}$\\
$^{1}$Center for Astrophysics and Space Astronomy, University of
Colorado, Boulder, CO 80309-0389; raymond@lasp.colorado.edu\\
$^{2}$Lunar and Planetary Laboratory, University of Arizona, Tucson, AZ 85721\\
$^{3}$NASA Goddard Space Flight Center, Greenbelt, MD 20771\\
$^{\dagger}$NASA Postdoctoral Program Fellow\\
$^{\star}$Member of NASA Astrobiology Institute\\
}

\date{Submitted Sept 20}

\pagerange{\pageref{firstpage}--\pageref{lastpage}} \pubyear{2002}

\maketitle

\label{firstpage}

\begin{abstract}

To date, two planetary systems have been discovered with close-in,
terrestrial-mass planets ($\la 5-10 \mearth$).  Many more such discoveries are
anticipated in the coming years with radial velocity and transit searches.
Here we investigate the different mechanisms that could form ``hot Earths''
and their observable predictions.  Models include: 1) {\it in situ} accretion;
2) formation at larger orbital distance followed by inward ``type 1''
migration; 3) formation from material being ``shepherded'' inward by a
migrating gas giant planet; 4) formation from material being shepherded by
moving secular resonances during dispersal of the protoplanetary disk; 5)
tidal circularization of eccentric terrestrial planets with close-in
perihelion distances; and 6) photo-evaporative mass loss of a close-in giant
planet.  Models 1-4 have been validated in previous work.  We show that tidal
circularization can form hot Earths, but only for relatively massive planets
($\ga 5 \mearth$) with very close-in perihelion distances ($\la$ 0.025 AU),
and even then the net inward movement in orbital distance is at most only
0.1-0.15 AU.  For planets of less than $\sim 70 \mearth$, photo-evaporation
can remove the planet's envelope and leave behind the solid core on a Gyr
timescale, but only for planets inside 0.025-0.05 AU.  Using two quantities
that are observable by current and upcoming missions, we show that these
models each produce unique signatures, and can be observationally
distinguished.  These observables are the planetary system architecture
(detectable with radial velocities, transits and transit-timing) and the bulk
composition of transiting close-in terrestrial planets (measured by transits
via the planet's radius).

\end{abstract}

%\keywords{planetary systems: formation --- planetary systems: observation --- extra-solar planets ---methods: n-body simulations --- astrobiology}

\section{Introduction}

Both radial velocity (RV) and transit searches are biased toward finding
large/massive planets at small orbital distances (e.g., Marcy \& Butler 1998;
Charbonneau \etal 2007).  Given the increased sensitivity of new instruments,
ever-smaller close-in planets are being detected.  Currently, two systems are
thought to contain close-in planets of less than 10 Earth masses ($\mearth$):
GJ 876 (Rivera \etal 2005) and GJ 581 (Udry \etal 2007).  Transit missions
{\it CoRoT} (Baglin 2003; Aigrain \etal 2007) and {\it Kepler} (Basri \etal 2005) expect to
find perhaps a few hundred close-in planets with masses less than 5-10
$\mearth$.  In this paper we focus on these ``hot Earth'' planets, which we
assume to have masses in the range $0.1 < m_p < 10 M_\oplus$, and semi-major
axes $a \la 0.2$ AU.

We propose that it is possible to determine the formation history of a given
hot Earth planetary system with two observable quantities: the architecture of
the inner planetary system, and the bulk composition of the hot Earth(s).
The planetary system architecture can be detected by a combination of RV and
transit measurements, as well as additional analysis of transit signals (e.g.,
timing variations: TTV; Agol \etal 2005, Holman \& Murray 2005).  The
composition of a transiting terrestrial planet can be determined by its
physical size, i.e. the transit depth.  Structure models indicate that very
water-rich planets ($\ga$10\% water by mass) have detectably larger radii
than dry, rocky planets or iron-dominated planets (Valencia \etal 2007a,
2007b; Fortney \etal 2007; Sotin \etal 2007; Seager \etal 2007), although a massive H/He envelope can also inflate the observed planetary radius (Adams \etal 2007).  

Several mechanisms for the formation of close-in terrestrial planets have been
proposed (Zhou \etal 2005; Gaidos \etal 2007).  In Section 2 we describe the
observable quantities that can distinguish between models.  In Section 3, we
summarize four known models, and test two unproven models: a) tidal
circularization of terrestrial planets on eccentric orbits, and b)
photo-evaporation of hot Neptunes or hot Jupiters.  We have tried to include
all reasonable models, which include various combinations of accretionary
growth, planet migration, and evaporative loss.  Table 1 summarizes the
observable differences between these models.  In section 4, we apply these
models to the two known hot Earth systems.  Section 5 concludes the paper with
a discussion of whether the mechanism for giant planet formation --
core-accretion or gravitational instability -- can affect the abundance of hot
Earths, as claimed by Zhou \etal (2005).

\section {Observable Quantities}

The observables considered in this paper are the architecture of the inner
planetary system and the bulk planetary composition.  The planetary system
architecture, i.e. the co-existence (or lack) of additional planets in hot
Earth systems, can provide strong circumstantial evidence for or against
certain formation models, as described below.  In particular, certain
characteristic planetary configurations are smoking guns (see Table 1).
Determining the bulk composition of a planet requires transit measurements.
Thus, our analysis applies only to systems with at least one transiting
planet.  In most cases, but not all, the transiting planet must be a hot
Earth.

The architecture of hot Earth planetary systems may be determined in three
primary ways: ($i$) via the detection of transits of multiple planets; ($ii$)
via radial velocity (RV) monitoring of the host star; and ($iii$) by analysis
of transit timing variations (TTV).  Other techniques such as astrometry may
be used in conjunction with these techniques, but note that astrometry is not
optimal for detecting close-in planets (Black \& Scargle 1982).  Detection of
multiple transiting planets in the same system requires extremely low mutual
inclinations between planetary orbits, which are thought to be rare (e.g.,
Levison \etal 1998).  The RV technique has discovered several planets with
minimum masses less than Neptune, including the two known systems with hot
Earths (Rivera \etal 2005; Udry \etal 2007).  There exist several
currently-operational instruments capable of RV followup for {\it CoRoT} and
{\it Kepler} targets, such as Keck HIRES (Vogt \etal 1994), the Hobby-Eberly
Telescope's HRS spectrograph (Cochran \etal 2004), and the HARPS instrument at
ESO (Mayor \etal 2003).  In addition, the HARPS-North spectrograph is being
built specifically to do RV followup of {\it Kepler} candidate transiting
planets (Latham 2007).  However, given that many of the target stars will be
very faint, RV followup of a large number of stars may not be possible.  For
those that can be followed up, the $\la m \, s^{-1}$ sensitivity of current RV
instruments should be able to detect close-in, $\la 5-10 \mearth$ planets and
to probe the inner regions of {\it CoRoT} and {\it Kepler}-detected targets.

%\begin{deluxetable}{p{4.5cm}|p{5cm}|p{5cm}}
%\tablewidth{0pt}
%\tablecaption{Observable Predictions of Hot Earth Formation Models}
%\tabletypesize{\normalsize}
%\tablecolumns{5}
%\renewcommand{\arraystretch}{.6}
%\tablehead{
%\colhead{Model} &  
%\colhead{System Architecture} & 
%\colhead{Planet Composition}}
%\startdata

\begin{table*}
 \centering
 \begin{minipage}{150mm}
  \caption{Observable Predictions of Hot Earth Formation Models}
  \begin{tabular}{p{5cm} || p{5cm} | p{5cm}}   %{@{}llrrrrlrlr@{}}
  \hline
   \bf \large Model & \bf \large System Architecture & \bf \large Planet Composition\\

\\ \hline

\bf {\it \bf In Situ} Accretion & Several hot Earths, spaced by $\sim$20-60 mutual Hill
radii & Relatively dry for Solar-type stars.  Up to 0.1-1 percent water for
low-mass stars \\
\hline

\bf Type 1 Migration & Chain of many terrestrial planets, close to mutual mean
motion resonances & Icy or Rocky, depending on formation zone.  Most likely to be icy ($\ga$ 10 \%
water by mass) . \\
\hline 

\bf Giant Planet Migration Shepherding & Co-existence of hot Earths and close-in
giant planets near (but not in) strong mean motion resonances & Rocky with
moderate water content: a few percent water by mass at time of formation.
\\ \hline 

\bf Secular Resonance Shepherding during Disk Dissipation & Co-existence of hot
Earths and at least two, interacting giant planets & Depends on the
details of the giant planet's orbital history.  Rocky if formed mainly {\it in situ}
\\   \hline

\bf Tidal Circularization of Eccentric Planets & Single hot Earth, with possible
distant companion (giant planet or stellar binary) to explain high
eccentricity & Depends on formation zone of planet -- rocky unless migrated
inward.
\\ \hline 

\bf Photo-evaporation of hot Neptunes & Hot Earth inside 0.025-0.05 AU.  Likely
chain of several planets, as for type 1 migration.  Correlation between hot
Earth vs. hot Neptune frequency and stellar age.  & Icy, assuming what remains
is a giant planet core.\\ \hline

\end{tabular}
\end{minipage}
\end{table*}

%\enddata
%\end{deluxetable}

Transit timing variations (TTV) analysis measures the deviation of a series of
transits from a perfect chronometer, representing a deviation of the
transiting planet's orbit from a perfect Keplerian ellipse due to
perturbations from one or more additional planets (Holman \& Murray 2005; Agol
\etal 2005).  The TTV signal scales with the transiting planet's orbital
period, and increases for more massive and closer perturbing planets.  For sufficiently accurate transit timing
data, TTV analysis can either derive the mass and orbit of a perturbing planet
or place constraints on the existence of nearby perturbers (Steffen \& Agol
2005; Agol \& Steffen 2007).  TTV is especially sensitive to planets that lie
in or close to mean motion resonances with the transiting planet, which is
convenient given that several formation models predict near-resonant planetary
configurations (see $\S$ 3 below).

The bulk composition of a planet determines its density and therefore its
physical size: ice-planets are far larger than iron-planets.  Recently,
several studies have calculated mass-radius relations for planets with
different compositions (Valencia \etal 2006, 2007a, 2007b; Fortney \etal 2007;
Sotin \etal 2007; Seager \etal 2007).  For a fixed mass, there exists a
roughly 40\% difference in radius between pure ice planets and pure rock
planets, and a similar 40\% difference between pure rock and pure iron
planets; these ratios of sizes are independent of planet mass.  In addition,
there is a $\sim$ 35\% difference in size between Earth-like planets (2/3
rock, 1/3 iron) and ocean planets (1/2 rock, 1/2 water; Fortney \etal 2007).\footnote{Note that water contents of a few percent by mass, though $\ga$10-20 times
larger than the Earth's estimated water budget (L\'ecuyer \etal 1998), would
have a negligible effect on the planetary radius compared with the
observational uncertainties.}  

Estimates of both the planetary mass and radius are needed to derive a bulk composition (Selsis \etal 2007).  For transiting planets, errors in stellar masses and radii (Ford \etal 1999;
Cody \& Sasselov 2002; Fischer \& Valenti 2005; Sozzetti \etal 2007) are
likely to lead to errors in planetary radii on the order of 2-10\% (see
section 6 of Seager \etal 2007).  With a determination of the planetary mass
to within 5\% and radius to within 10\%, it may be possible to differentiate
between mostly rocky (Earth-like) planets and icy planets with $\geq$ 10\%
water by mass (Valencia \etal 2007b).  For transiting planets with very precise
mass and radius measurements (better than 2\%), more detailed compositions may
be derived (Seager \etal 2007).  

Atmospheric envelops of H/He can inflate the observed radii of solid planets by tens of percent (Adams \etal 2007).  For a given radius measurement, solutions for the planetary structure become degenerate with respect to water content and envelope mass: small radii are clear signatures of rocky planets, but larger radii are ambiguous.  Theoretical models suggest that the ability of a planet to accrete a gaseous envelope depends on the planet's mass, and is not sensitive to the orbital distance (e.g., Ikoma \etal 2001; Ida \& Lin 2004).\footnote{In addition, as Adams \etal (2007) point out, significant envelopes of hydrogen may be formed as a result of outgassing from the planetary interior.}  Thus, additional information about the presence and thickness of the planet's atmosphere is needed to determine whether the planet is water-rich or rocky.   For example, more information could be gathered from a situation in which an atmosphere has almost certainly photo-evaporated away (old stellar age plus very close-in planet).  Another favorable case would be a situation for which some spectral information about the planet's atmosphere could potentially be obtained.  

Observational limitations are such that certain planets will have no mass estimates, given the faintness of their host stars and the consequent difficulty of RV followup.  For such cases, it is possible to place
mass limits based on maximum and minimum radius estimates, i.e., by assuming
the planet to be made of pure iron or pure water (or pure hydrogen for gaseous
planets).  In addition, the bulk planetary composition may not be determined in many cases because of observational limitations (Selsis \etal 2007) or degeneracy between model parameters (Adams \etal 2007).  With no composition information it becomes more difficult
to differentiate between formation models.  Nonetheless, several cases can be
distinguished if the inner planetary system architecture is known.  

Thus, current and future programs have the sensitivity to determine the
orbits, masses, radii and companions of a large number of hot Earths.
Although it will not be feasible in all cases, information about other planets
in hot Earth systems will be determined via RV, transits, and transit timing
analysis.  In this paper we focus on systems in which both the inner planetary architecture and the composition of a hot Earth (rocky vs. $>$10\% water) can be determined (i.e., the brightest {\it CoRoT} and {\it Kepler} targets; see FIg.~6 of Selsis \etal 2007). 
As explained below and summarized in Table 1, analysis of these
data may be able to identify the formation mechanism of such planets.

\section {Models for hot Earth formation}

Here we investigate six mechanisms for hot Earth formation, including proven and
previously untested mechanisms.  The six models are: ($\S3.1$) {\it in situ} accretion; ($\S3.2$) type 1 migration; ($\S3.3$) shepherding during giant planet migration; ($\S3.4$) shepherding via secular resonance sweeping; ($\S3.5$) tidal circularization of eccentric planets; and ($\S3.6$) photo-evaporation of close-in giant planets.  For each model, we discuss the state of the inner planetary system, as well as the likely composition of the hot Earth(s). Table 1 summarizes the differences between models.  The first four models listed have been demonstrated in previous work.  We introduce two additional models for hot Earth formation, and test them quantitatively below.

\subsection{{\it In situ} formation}  

If protoplanetary disks contain a substantial mass in solids close to their
stars, then perhaps hot Earths can form from local material.  This depends
critically on the condensation temperatures of grains (Pollack \etal 1994;
Lodders 2003), disks' inner truncation radii (e.g., Eisner \etal 2005; Akeson
\etal 2005), and the surface density profile of solids (Weidenschilling
1977; Hayashi 1981; Davis 2005; Raymond \etal 2005).  If hot Earths form {\it
in situ}, then their growth would be similar to that of Solar System's
terrestrial planets (Wetherill 1990, 1996; Chambers \& Wetherill 1998; Agnor
\etal 1999; Morbidelli \etal 2000; Chambers 2001; Kenyon \& Bromley 2006), but minus the dynamical effects of Jupiter and Saturn (although giant planets may co-exist with some hot Earths, e.g. Gliese 876; Rivera \etal 2005).

If there is sufficient mass to form one hot Earth {\it in situ}, then we
expect a population of several hot Earths to form, with masses determined by
the local disk mass and spacings similar to those in the Solar System and in
accretion simulations (roughly 20-80 mutual Hill radii: $R_{H,m} = 0.5 (a_1 + a_2)
(M_1+M_2/3 M_\star)^{1/3}$; $a_1$ and $a_2$ are the orbital radii and $M_1$
and $M_2$ the masses of two adjacent planets).  The surface density
distribution of protoplanetary disks, $\Sigma$, is thought to scale with
radial distance $r$ as $\Sigma \sim f r^{-\alpha}$, where $f$ is a scale
factor to account for the large variability in observed disk masses (Andre \&
Montmerle 1994; Eisner \& Carpenter 2003; Andrews \& Williams 2005; Scholz
\etal 2006), and the value of $\alpha$ lies between 0.5 and 2 (Weidenschilling
1977; Hayashi 1981; Kuchner 2004; Davis 2005; Dullemond \etal 2007; Andrews \&
Williams 2007; Garaud \& Lin 2007).  Accretion models suggest that planets
close to their stars are generally smaller than those farther out for $\alpha
< 2$ (Lissauer 1987; Kokubo \& Ida 2002; Raymond \etal 2005, 2007; Kokubo
\etal 2006).

For solar-type stars, hot Earths that form {\it in situ} are likely to be dry
because of the low efficiency of water delivery from both comets (Levison
\etal 2000) and asteroids (Raymond \etal 2004).  Because of the very hot local
temperatures, these planets would be mainly composed of refractory materials
such as iron and rock (Pollack \etal 1994; Lodders 2003).  However, for the
case of low-mass stars, the snow line is located very close-in (as is the
habitable zone -- Kasting \etal 1993).  Water delivery to hot Earths may
therefore be more favorable around low-mass stars, although impact speeds are
high and formation times fast compared with Earth's formation zone (Lissauer
2007; Raymond \etal 2007), and the snow line moves significantly in the disk
lifetime (Sasselov \& Lecar 2000; Kennedy \etal 2006).

\begin{figure*}
  \begin{center} \leavevmode \epsfxsize=15cm\epsfbox{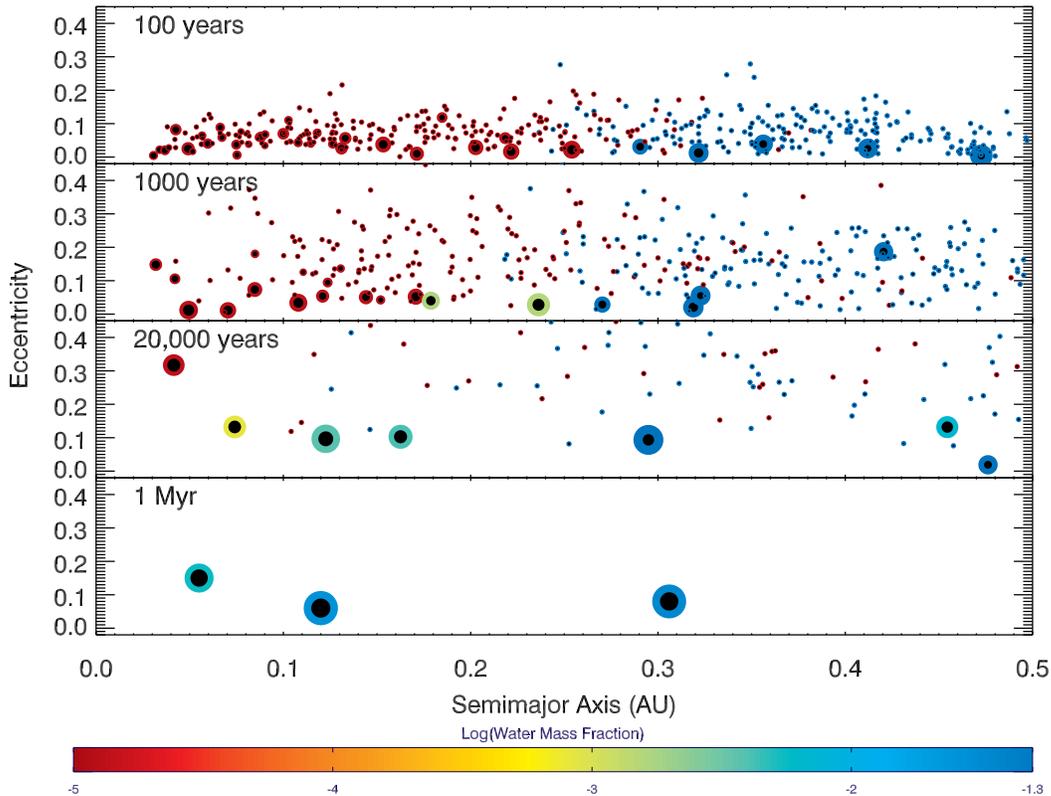}
    \caption[]{Snapshots of {\it in situ} accretion of a system of hot Earths.
    Each object's size is proportional to its mass$^{1/3}$, and the color
    corresponds to its water content, from 10$^{-5}$ water by mass (red) to
    5\% water by mass (darkest blue; see color bar).  The dark circle in the
    center of each body refers to the relative size of its iron core (see
    Raymond \etal (2005) for details).  In this case, the central star is 0.31
    $\msun$, the same as Gliese 581.}  \label{fig:insitu} \end{center}
\end{figure*}

Figure~\ref{fig:insitu} shows snapshots of {\it in situ} accretion of
terrestrial material close to a 0.31 $\msun$ star from Raymond \etal (2007, in
preparation), designed to examine the GJ 581 system.  The simulation started
from a disk of 57 planetary embryos (initially separated by 3-6 Hill radii, as
in Raymond \etal 2006a) and 500 planetesimals in a very massive disk totaling
40 $\mearth$ between 0.03 and 0.5 AU.  The disk's surface density decreased
with orbital distance $r$ as $r^{-1}$ (i.e., $\alpha = 1$), and was roughly 30
times more massive than the minimum-mass solar nebula model (Weidenschilling
1977; Hayashi 1981; Davis 2005).\footnote{Note that this disks as massive as the one in Fig.~\ref{fig:insitu} (30 times the minimum-mass disk) are likely to be quite rare (e.g., Andrews \& Williams 2005).  Thus, {\it in situ} accretion of close-in planets of several Earth masses can probably occur only for a small fraction of protoplanetary disks (see $\S$ 4).}  The three planets that formed in this
simulation have masses of 6.6 $\mearth$ (at 0.06 AU), 10.9 $\mearth$ (0.12
AU), and 10.6 $\mearth$ (0.30 AU).  Each has a substantial water content,
acquired via collisions with material originating beyond the "water line" at
0.29 AU, but note that the effects of water depletion during impacts (Genda \&
Abe 2005; Canup \& Pierazzo 2006) and hydrodynamic escape (Matsui \& Abe 1986;
Kasting 1988) have not been accounted for.

Thus, if hot Earths form {\it in situ} then we expect systems of several hot
Earths to form, with spacings comparable to the solar system terrestrial
planets.  These planets will contain mainly local, dry material, although for
low-mass stars they may contain up to perhaps 0.1-1\% water by mass (but see
Raymond \etal 2007, Lissauer 2007).

\subsection {Inward Type 1 Migration}  

Earth- to $>$Neptune-mass planets excite density waves in the disk
(Goldreich \& Tremaine 1979).  The back reaction of these waves on the planet
causes inward orbital migration on a timescale of 10$^{4-6}$ years (Goldreich
\& Tremaine 1980; Ward 1986, 1997; D'Angelo \etal 2003; Masset \etal 2006a).
However, type 1 migration may be stopped or even reversed in the very inner
regions of the disk, because of net disk torque changes at disk edges (Masset
\etal 2006b) or in the optically thick inner disk (Paardekooper \& Mellema
2006).  Thus, hot Earths could form far from their stars and migrate in to the
location in the disk where the disk torques cancel out and migration is
stopped.  Presumably, if there is enough solid and gas mass to enable type 1
migration of one planet, then others should follow.  Co-migrating planets may
end up trapped in mean motion resonances (e.g., Lee \& Peale 2002), and can
form chains of many planets in paired resonances.  These orbital chains of
planets can survive for long times, as the outward-directed torques on the
inner planets may be balanced by inward-directed torques on the outer planets
(Terquem \& Papaloizou 2007).  Surviving planets do not remain on strictly
resonant orbits, and collisions between planets can occur after the disk
dissipates.

The amount of solid material in the disk is thought to increase by a factor of
2-4 or perhaps more beyond the snow line (Hayashi 1981, Stevenson \& Lunine
1988; Lodders 2003).  In addition, most disk surface density profiles contain
far more mass in their outer regions.  Thus, it seems reasonable to assume
that, in this model, most hot Earths must form beyond the snow line and are
therefore icy in composition rather than rocky.  Indeed, transits of the hot
Neptune GJ 436 b have been interpreted as an indication that it may be largely composed of water and
may therefore have formed beyond the snow line and migrated inward (Gillon \etal 2007).
Note, however, that inferring a detailed planetary composition from a radius measurement is ambiguous because different combinations of rock, ice and H/He envelopes can form planets with the same mass and radius (see Fig.~3 from Adams \etal 2007).  In addition, formation and migration models suggest that in a system of several hot Earths it is possible for the innermost hot Earth to be rocky (Alibert \etal 2006).

The main consequence of the type 1 migration model is simply that hot Earths
didn't form locally but farther out in the disk, probably in the
water-rich icy regions.  Therefore, hot Earths should contain a large quantity
of ice and have measurably larger radii.  In addition, the migration process
favors the formation of a chain of resonant or near-resonant planets (Terquem
\& Papaloizou 2007).

\subsection{Shepherding by giant planet migration}

Giant planets more massive than a critical value carve an annular gap in the
protoplanetary disk and are thus locked to the disk's viscous evolution (Lin
\& Papaloizou 1986; Takeuchi \etal 1996; Bryden \etal 1999; Rafikov 2002;
Crida \etal 2006).  These planet subsequently ``type 2'' migrate, usually
inward, on a $\sim 10^5 - 10^6$ year timescale, depending on the disk's
viscosity (Lin \& Papaloizou 1986, Lin \etal 1996; Ward 1997; D'Angelo \etal
2003).  Such a planet migrates through a disk composed of both gas and solids
in the form of km-sized planetesimals and Moon- to Mars-sized planetary
embryos, which formed in series of dynamical steps from micron-sized dust
grains (as in the {\it in situ} formation model; see $\S$ 3.1 or Chambers
2004, Papaloizou \& Terquem 2006 for reviews).  As the giant planet migrates
inward, it shepherds material in front of strong mean motion resonances
(MMRs).  The evolution of a typical planetary embryo in the inner disk
proceeds as follows.  As the giant planet approaches the embryo, the embryo's
eccentricity is increased by an MMR (usually the 2:1 or 3:2, but higher-order
resonances are stronger for more eccentric giant planets; Murray \& Dermott
1999).  Gas drag and dynamical friction with nearby planetesimals act to
recircularize the embryo's orbit and decrease its energy, thereby reducing its
semimajor axis and moving it just interior to the MMR (Adachi \etal 1976;
Tanaka \& Ida 1999).  As the giant planet continues its migration, the embryo
is again excited by the approaching MMR and the cycle continues.  Thus,
embryos and planetesimals are shepherded inward by moving MMRs and accrete
into planet-sized bodies during giant planet migration\footnote{The formation
timescale of shepherded hot Earths is on the order of the migration timescale
(Mandell \etal 2007).  Thus, hot Earths may form in $\sim 10^{5-6}$ years, as
opposed to the 10$^{7-8}$ year timescale for the Earth calculated from Hf/W
isotopic measurements (Kleine \etal 2002; Jacobsen 2005).  This very short
formation timescale for hot Earths could have consequences for their
geological evolution.}  (Fogg \& Nelson 2005, 2007; Zhou \etal 2005; Raymond
\etal 2006b; Mandell \etal 2007).  However, during this process, many bodies'
eccentricities are damped too slowly to avoid a close encounter with the giant
planet.  Such bodies are usually scattered outward, and can form a subsequent
generation of exterior terrestrial planets (Raymond \etal 2006b; Mandell \etal
2007). 

\begin{figure*}
  \begin{center} \leavevmode \epsfxsize=15cm\epsfbox{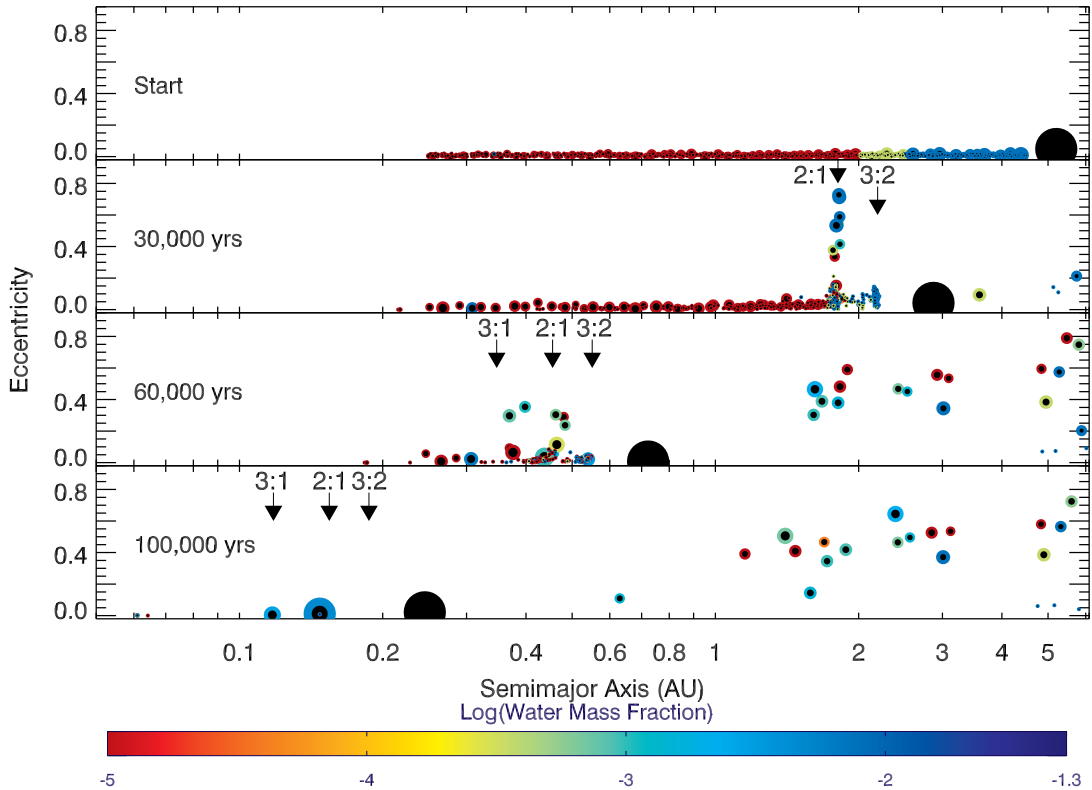}
    \caption[]{Snapshots in time of the migration of a Jupiter-mass giant
    planet through a disk of terrestrial bodies, including the effects of
    aerodynamic gas drag, from a simulation by Mandell \etal (2007).  Specific
    strong resonances with the giant planet are indicated.  Colors refer to
    bodies' water contents, as in Fig.~\ref{fig:insitu}.  This version
    reproduced from Gaidos \etal (2007).}  \label{fig:migshep} \end{center}
\end{figure*}

Figure~\ref{fig:migshep} shows snapshots in time of this shepherding process
from a simulation by Mandell \etal (2007).  It is clear that the 2:1 MMR is
responsible for the bulk of the shepherding in this simulation.  The two hot
Earths formed are on low-eccentricity orbits immediately interior to strong
resonances, as expected.  However, in simulations including weaker gas drag,
hot Earths can form on higher eccentricity orbits (Fogg \& Nelson 2007,
Mandell \etal 2007).  The survival of high eccentricity hot Earths is
uncertain, given that tides may act to alter the planets' orbits and possibly
lead them into unstable giant planet resonances or drive them into the star
(see $\S$ 3.5 below).

In the giant planet migration shepherding model, hot Earths are a mixture of
material that originated interior to the giant planet's orbit.  Both the
core-accretion and gravitational collapse models predict that giant planets
are likely to form at large orbital distances, beyond the snow line, which
itself moves inward in time (Pollack \etal 1996; Boss 1997; Bodenheimer \etal
2000; Sasselov \& Lecar 2000; Mayer \etal 2002).  In the simulations of
Raymond \etal (2006b) and Mandell \etal (2007), the hot Earths that formed
contained 1-2\% water by mass, as is the case for the two hot Earths in
Fig.~\ref{fig:migshep}.  The assumed starting water distribution in those
cases was similar to that of current-day primitive asteroids (Abe \etal 2000;
Fig.~2 from Raymond \etal 2004).  Note that water depletion from impacts
and hydrodynamic escape was neglected in these calculations.

The giant planet migration shepherding model predicts that hot Earths
should lie close to a strong MMR, most likely the 2:1 MMR, interior to a giant
planet (Fogg \& Nelson 2005; Zhou \etal 2005; Raymond \etal 2006b; Mandell
\etal 2007).  In this model, hot Earths are formed from a mixture of material
that originated interior to the giant planet's orbit.  Giant planets are
expected to form just outside the snow line, given the increase in solid
material (Hayashi 1981; Stevenson \& Lunine 1988; Ida \& Lin 2004).  Thus, hot
Earths formed by shepherding are likely to contain up to a few percent water,
but probably not more (Mandell \etal 2007).  Note that planets with water
contents of a few percent by mass cannot be distinguished from rocky planets
by radius measurements given observational uncertainties (Valencia \etal
2007b; Fortney \etal 2007; Sotin \etal 2007; Seager \etal 2007).

\subsection{Shepherding by Sweeping Secular Resonances during Disk Dispersal}

Moving secular resonances (SRs) can shepherd material in a similar way to mean
motion resonances (MMRs) if gas drag is present.  An SR occurs when the
apsidal precession frequency of two bodies' orbits are commensurate (e.g.,
Murray \& Dermott 1999).  In a disk with two or more giant planets,
interactions between the planets cause each of their orbital alignments to
precess.  In addition, the gravitational potential of the massive gaseous disk
affects the precession rates, and therefore the location of SRs with each
planet in the disk (Ward 1981; Nagasawa \etal 2005).  As the disk dissipates,
SRs can move progressively (``sweep'') across a given region, increasing the
eccentricities of bodies.  In the case of a smooth, inward-sweeping SR,
shepherding of material can happen similar to MMR shepherding for migrating
giant planets. 

In the context of hot Earth formation, the SR shepherding model applies to
cases with two or more giant planets that have stopped migrating.  A smooth
dissipation of the disk can induce SR sweeping.  Much as in the migration
shepherding mechanisms, a sweeping SR excites the eccentricities of nearby
protoplanets.  These eccentricities are subsequently damped by gas drag and
the body's orbit is moved interior to the resonance.  This process continues
for the duration of the SR sweeping, unless a planet gets close enough to the
star that its precession rate becomes dominated by general relativistic
effects rather than dynamical ones (Zhou \etal 2005).

Thus, the secular resonance shepherding model involves a complex interaction
between two giant planets, the massive gaseous disk, and relatively low-mass
terrestrial material.  It requires a monotonic, inward secular resonance
sweeping which itself requires a smooth dispersal of the gaseous disk (Ward
1981), which is uncertain given that most stars form in large clusters and may
lose disk mass in periodic photo-evaporation events (Lada \& Lada 2003; Adams
\etal 2004; Hester \etal 2004).  In the SR shepherding model, a hot Earth
system must also contain at least two more distant, interacting giant planets.
The compositions of hot Earths in this scenario are a mixture of material from
interior to the giant planets' starting orbits.  Estimating the compositions
of hot Earths in this model therefore requires a knowledge of the giant
planet's formation locations, specifically how far past the snow line they
formed.

\subsection{Tidal Circularization of Eccentric Terrestrial Planets}

The circular orbits of hot Jupiters have been attributed to energy and angular
momentum dissipation via tides raised on the planet by the star (Rasio \etal
1996).  In fact, it has been proposed that close-in giant planets may have
been scattered onto high-eccentricity orbits and tidally circularized (Rasio
\& Ford 1996; Weidenschilling \& Marzari 1996; Mardling \& Lin 2004; Jackson
\etal 2007).  Could tidal circularization act as a mechanism to transport
terrestrial planets inward?  To address this possibility, we integrated the
second-order, coupled semimajor axis $a$ and eccentricity $e$ tidal evolution
equations (Goldreich \& Soter 1966; Kaula 1964; Greenberg 1977):
\begin{equation} 
\frac{da}{dt} =
-\Big(21\frac{\sqrt{GM_*^3}R_p^5k_p}{m_pQ'_p}e^2
+
\frac{9}{2}\frac{\sqrt{G/M_\star}R_\star^5m_pk_\star}{Q'_\star}\Big)a^{-11/2}
\end{equation}

\begin{equation} 
\frac{de}{dt} =
-\Big(\frac{21}{2}\frac{\sqrt{GM_\star^3}R_p^5k_p}{m_pQ'_p}
+
\frac{171}{16}\frac{\sqrt{G/M_\star}R_\star^5m_pk_\star}{Q'_\star}\Big)a^{-13/2}e
\end{equation}

\noindent where $Q'_p$ and $Q'_\star$ are the tidal dissipation functions of
the planet and star, respectively, $k_p$ and $k_*$ are the Love numbers of the
planet and star, $m_p$ and $M_\star$ are the masses of the planet and star,
$R_p$ and $R_\star$ are the radii of the planet and star, and $G$ is the
gravitational constant.

Note that solutions with higher order terms in $e$ have been derived (e.g.,
Hut 1981; Eggleton \etal 1998). However, such models include, in effect,
assumptions about how a body responds to the ever-changing tidal potential,
effects that have not been observed. Therefore not enough is known about the
actual response of real bodies to evaluate these higher-order effects. As we
are only interested in the qualitative differences between planets whose
orbits have evolved through tidal decay and those that did not, the second
order solution should suffice.

We considered a stellar mass of 0.3, 1, and 3 $\msun$ with radii determined
from Gorda \& Svechnikov (1999), a planet mass of 1 and 5 $\mearth$ (assuming
$R_p \propto m_p^{0.27}$, as suggested by Valencia \etal 2006), a perihelion
distance from 0.025 to 0.1 AU, and eccentricities $e$ from 0 to 0.9.  For the
planet, we assumed $k_p = 0.3$ and $Q'_p = 21.5$ (Dickey
\etal 1994; Mardling \& Lin 2004); for the star, $k_\star = 1.5$ and 
$Q'_\star = 10^{5.5}$ (Ogilvie \& Lin 2007; Jackson \etal 2007). Each orbit
was integrated for 10 Gyr using a 10$^3$ year timestep, which convergence
tests showed is three orders of magnitude smaller than necessary to produce
reliable results.

\begin{figure*}
  \begin{center} \leavevmode \epsfxsize=15cm\epsfbox{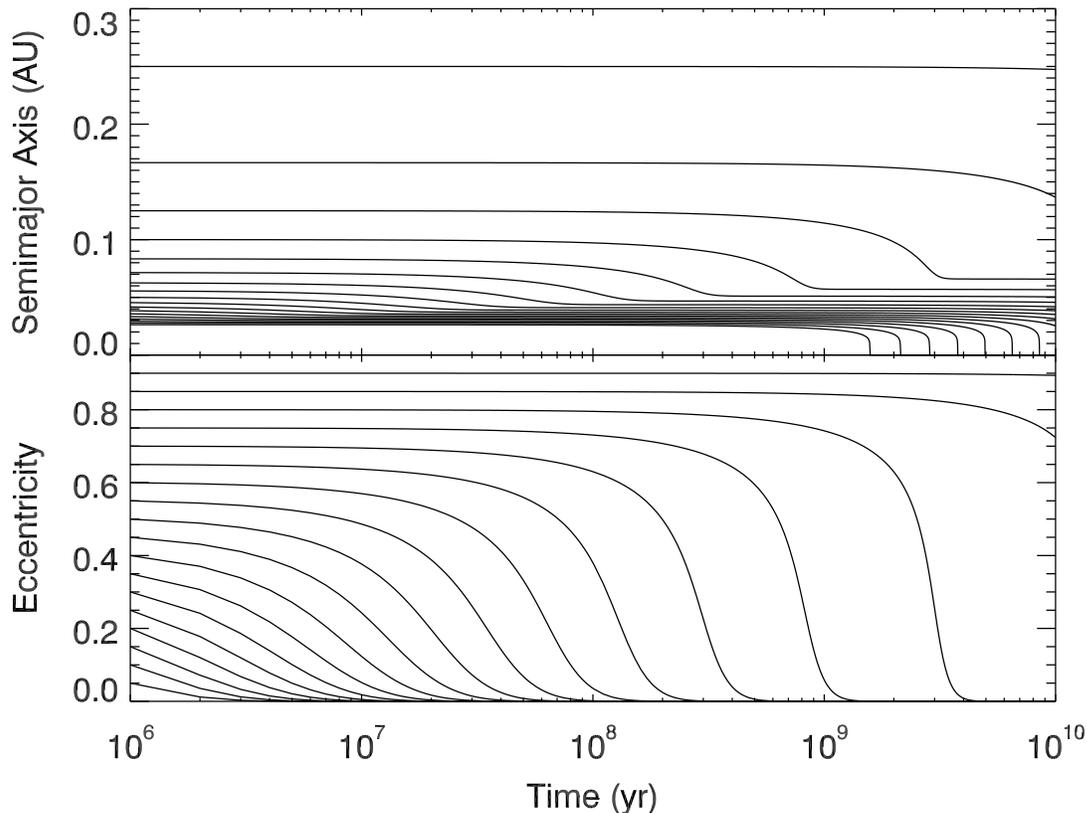}
    \caption[]{Orbital semimajor axis $a$ and eccentricity $e$ vs. time for a
    series of $5 \mearth$ planets orbiting Sun-like stars with starting
    perihelion distances of 0.025 AU.  Each curve corresponds to a planet with
    a given starting eccentricity, from 0.05 ($a$ = 0.026 AU) to 0.9 ($a$ =
    0.25 AU). If $a$ drops below $\sim 0.01$ AU, then the planet will fall
    into the star due to tidal evolution.}  \label{fig:tides} \end{center}
\end{figure*}

Figure~\ref{fig:tides} shows the evolution of a set of 5 $\mearth$ planets
with the same initial perihelion distance of 0.025 AU, and starting
eccentricities ranging from 0.05 ($a$ = 0.026 AU) to 0.9 ($a$ = 0.25 AU).  As
expected, evolution proceeds much faster for bodies at smaller orbital
distances (in this case, those with lower eccentricities).  For planets with starting eccentricities of 0.6 or less ($a \leq 0.06$ AU), orbital circularization takes place within $10^8$ years, including
an inward drift in semimajor axis of up to 0.01-0.02 AU.  Circularization
takes longer for larger $a$ values, but the amount of inward drift is also
increased.  For the $e$ = 0.8 planet, circularization requires several Gyr,
but the planet moves inward from 0.125 to 0.065 AU.  At still-larger orbital
distances (and eccentricities), circularization takes longer than the age of
the star.  Note that orbital evolution continues slowly after the planet's
orbit becomes circular, via tides raised on the star by the planet.  Mardling
\& Lin (2004) showed that should $a$ become very small ($\la 0.01$ AU),
then the planet is doomed to fall into the star within a few Gyr. We
confirm that assessment here. Therefore we expect no planets inside 0.01 AU
for any formation scenario.

The degree to which tidal circularization can move a planet inward clearly
depends on its starting orbit, size, and ability to dissipate energy.
Figure~\ref{fig:tides} shows the most tidal evolution of any of the cases we
explored for a Solar-mass star; in most cases evolution was slower (except for
the 3 $\msun$ cases, which were faster).  Thus, it appears that inward
movement of planets during tidal evolution is relatively small, at most a
$\sim$0.05 AU change in semimajor axis.  However, if the large eccentricity
were due to a perturbative event, then the planet's starting aphelion might be
representative of its pre-encounter semimajor axis.  In that case, the
effective inward movement due to tides is doubled (starting aphelion to final
semimajor axis), although it would still be less than about 0.1-0.15 AU on a
$>$Gyr timescale.  The composition of the planet depends on its formation
history, especially whether it formed locally or migrated inward.

What is the source of the large eccentricity needed to drive tidal circularization?
Planet-planet scattering has been invoked to explain the large eccentricities
of the known extra solar planets (Rasio \& Ford 1996; Weidenschilling \&
Marzari 1996; Lin \& Ida 1997; Ford \etal 2005; but see Barnes \& Greenberg
2007).  The strength of a scattering event depends on a combination of the
escape speed of the perturber, the encounter velocity, and the escape speed
from the system.  For close-in planets, the system escape speed is large, and
so only very massive bodies can excite large eccentricities.  Indeed,
accretion may be preferred over scattering in these situations (Goldreich
\etal 2004).

One alternative mechanism for eccentricity growth is an instability in nearby
giant planets could alter the orbit of a terrestrial planet (Veras \& Armitage
2006).  In that case, one or more exterior giant planets should exist in the
system, on eccentric orbits.  Another possible source of eccentricity could
arise if the host star had a binary companion.  If the orbital plane of the
planet were significantly inclined with respect to that of the binary, then
large eccentricities could be induced via the Kozai mechanism (Kozai 1962).
In most of these models, some evidence for an external perturber should be
evident.

Thus, tidal circularization can move a highly-eccentric terrestrial planet
inward to some extent, although the planet must be relatively massive ($\ga 5
\mearth$) and have a very small starting perihelion distance ($\la$ 0.03 AU).
If the planet formed locally, then its composition is likely to be relatively
dry ($\la$ 1\% water by mass).  A source of high eccentricity may also be
evident, such as a binary stellar companion or a distant eccentric giant
planet.

\subsection{Giant Planet Migration and Photo-evaporation}

Baraffe \etal (2004, 2006) proposed that close-in, Neptune-mass planets might be the
remains of larger planets that have been photo-evaporated away.  Such planets
would form farther from their parent stars and migrate inward (Ida \etal 2004,
Alibert \etal 2005), losing a portion of their gaseous envelopes
hydrodynamically via irradiative XUV heating (Lammer \etal 2003, Baraffe \etal
2004).  Here we investigate the possibility that photo-evaporation could lead
to the removal of the entire envelope of a hot Jupiter or hot Neptune, leaving
behind a solid planet, i.e., the core of the irradiated giant planet.

Recent estimates derive evaporation rates that are far smaller than those used
by Lammer \etal (2003), and include the effects of
two-dimensional layering (Tian \etal 2005) and improved atmospheric chemistry
(Yelle 2004, 2006).  Indeed, these new evaporation rates are closer
to those of Watson \etal (1981).  Perhaps most convincing that the Lammer
\etal evaporation rates are too large is empirical evidence that the mass
distribution of highly irradiated extra-solar planets (inside 0.07 AU) is
identical to that of more distant planets (Hubbard \etal 2007a).  A
substantial change in the mass function is predicted for evaporation models
(i.e., fewer massive planets and more less-massive ones).\footnote{Fortney \etal (2007) also showed that if hot Neptunes form via photo-evaporation of hot Jupiters, then their radii should be on the order of one Jupiter radius.  However, if they are not remnants of hot Jupiters, then their radii should be 0.3-0.4 $R_J$.  The first transiting hot Neptune indeed has a radius of $\sim 0.35-0.4 R_J$ (Gillon \etal 2007; Deming \etal 2007).  Note, however, that Gliese 436 is an M dwarf (0.41 $\msun$) and therefore has low EUV and FUV emission (except during flares), which are key for driving evaporative mass loss (Butler \etal 2004).} Such an effect may exist at lower masses, but not in the currently-probed sample of planets.
Hubbard \etal (2007b) show that at the minimum orbital radius of known
extra-solar planets (0.023 AU), the initial mass must be less than about a
Saturn mass to evaporate completely, i.e., to its core.  For more typical hot
Jupiter orbits, at 0.05-0.1 AU, this critical mass is smaller
still.\footnote{Tidal evolution models suggest that close-in planets may have
originated at somewhat larger orbital distances and slowly evolved inward,
concurrently decreasing their semimajor axes and eccentricities (Jackson \etal
2007).  If close-in giant planets did indeed originate on somewhat more
distant, eccentric orbits, then their time-averaged fluxes would likely be
reduced.}  The models of Baraffe \etal (2004) and Hubbard \etal (2007b) do not account for the presence of a core, which is important once the planet mass is less than $\sim 100 \mearth$, such that a 5-10 $\mearth$ core constitutes a non-negligible fraction of the planet mass.  Note that the Baraffe \etal (2006) models do incorporate this effect, as we do implicitly by using their internal structure models.  

Mass loss due to hydrodynamic escape, limited only by energy deposition, depends critically on the stellar irradiance of the atmosphere, and can be approximated by the relation
\begin{equation}
\dot{M} = \frac{3\, \beta (F_* , a)^3}{G\rho}\, \frac{F_{XUV} + F_{\alpha}}{a^2}
\end{equation}
where $F_{XUV}$ and $F_{\alpha}$ represent the high-energy radiation incident on the
planet, and $\rho$ and $a$ are the planet density and orbital distance from
the star respectively (Lammer \etal 2003; Baraffe \etal 2004). The parameter
$\beta$ is the ratio of the irradiated planetary radius to the
planet's ``original'', non-irradiated radius for a specific stellar flux and orbital distance (Lammer \etal 2003, Baraffe \etal 2004, 2006, and Hubbard \etal 2007b all assume $\beta$ = 3 based on atmospheric models of Watson \etal 1981).  For a constant orbital distance, the mass loss will
therefore initially decrease in time as the planet cools and becomes more
dense and the star's UV and x-ray flux decreases.  Over time, as hot material
escapes from the top of the planetary atmosphere new layers are irradiated and
stellar flux is converted to expansion energy, gravitational contraction of
the planet slows.  If enough mass is evaporated, expansion surpasses
contraction and the planet experiences run-away mass loss, leaving behind only
the solid core.

To construct a simplified model of photo-evaporative mass loss, we need to
constrain certain parameters.  We assumed evaporation rates 100 times smaller
than the energy-limited case from Lammer \etal (2003).  The radius of an
evaporating planet stays relatively constant regardless of mass, such that we
extrapolated radii for planets of various masses and heavy-element abundances
from Baraffe \etal (2006, with corrections for semimajor axis from Chabrier
\etal 2004) to find a mass-radius relation for irradiated planets as a
function of time. Planets less massive than $\sim50 \mearth$ are more likely than larger planets to contain significant concentrations of molecular species, simply because the ratio of core mass to envelope mass is decreased (e.g., Uranus and Neptune).  Molecules such as $\rm H_2O$ and $\rm CH_4$ play an important part in the energetics of atmospheric expansion, and therefore affect the mass loss rate (Hubbard \etal 2007a).  Although our approach does not directly incorporate changes in evaporation rate with chemistry, the mass-radius relations from Baraffe \etal (2006) are based on Alibert \etal (2005)Õs values for heavy-element enrichment of the planets' atmospheres and therefore implicitly include a decreased evaporation rate for smaller planets with heavy-element-rich atmospheres since evaporation rates depend on the planet radius.  Our simplified model demonstrates good agreement with the results from more detailed models by Hubbard et al. (2007b) and the reduced-evaporation models of Baraffe et al. (2006) for higher-mass planets.

\begin{figure*}
  \begin{center} \leavevmode \epsfxsize=15cm\epsfbox{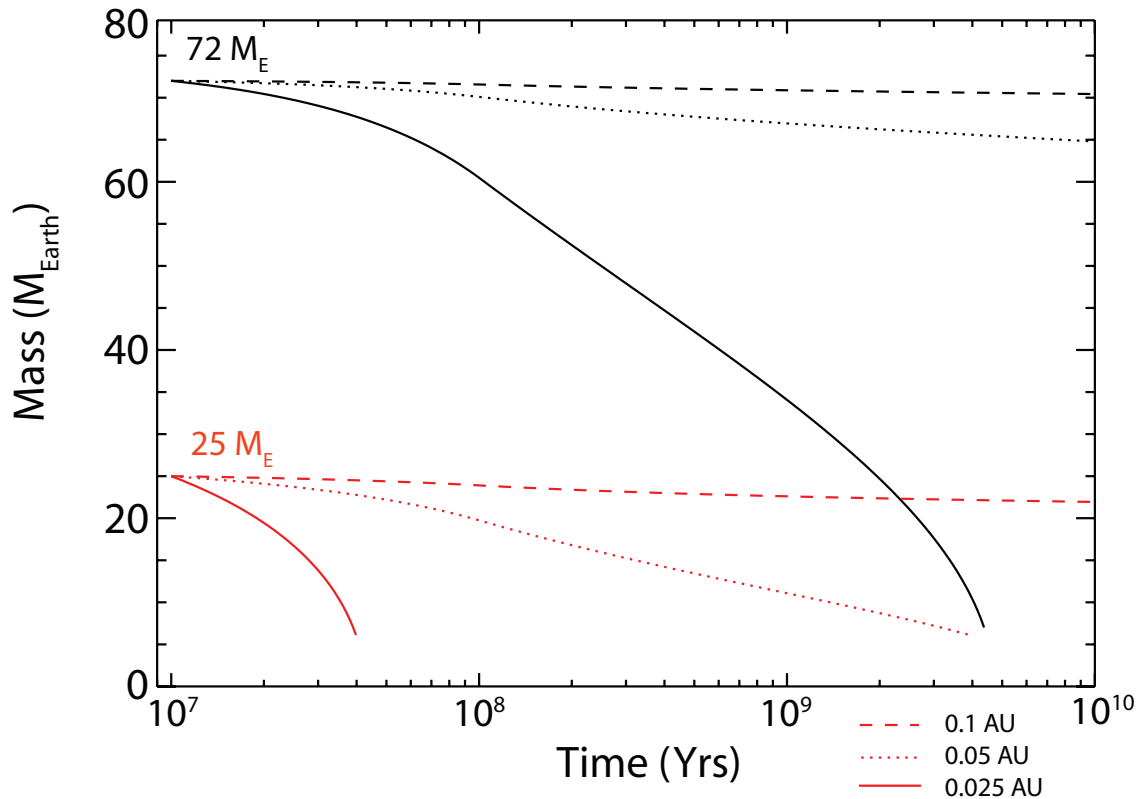} \caption[]{The
    evolution of the masses of two highly-irradiated planets due to
    photo-evaporation.  Our model is based on planetary structure models of
    irradiated planets from Baraffe \etal (2006) and Chabrier \etal (2004) and
    evaporation rates based on the model of energy-limited hydrodynamic escape
    from Lammer \etal (2003), but reduced by a factor of 100 in accordance
    with the results of Hubbard \etal (2007). The intermediate-mass planet
    ($72 \mearth$) represents the boundary between planets that would undergo
    type I migration (M$\la 70 \mearth$) and planets that would undergo type
    II migration (M$\ga 70 \mearth$; D'Angelo \etal 2003). The planets are
    placed at three different orbital radii: 0.025 AU (solid), 0.05 AU
    (dotted), and 0.1 AU (dashed).}  \label{fig:photo} \end{center}
\end{figure*}

Figure~\ref{fig:photo} shows the masses of two highly-irradiated planets as a
function of time, on circular orbits at 0.025, 0.05, and 0.1 AU.  Even on its
closest orbit, the more massive planet ($72 \mearth$) took 5 Gyr to evaporate
to its core.  Given that the transition from type 1 to type 2 migration is
thought to occur at roughly 70 $\mearth$ (Ward 1997; D'Angelo \etal 2003),
this effectively rules out the formation of hot Earths by type 2 migration and
subsequent photo-evaporation, simply because planets massive enough to type 2
migrate will not lose enough mass by photo-evaporation.  In contrast, the less
massive planet ($25 \mearth$) evaporated to its core in less than 50 Myr at
0.025 AU, but required Gyr to lose mass past 0.05 AU.  Additionally, beyond
$\sim$0.1 AU the mass loss over 5 Gyrs is negligible; planets found
beyond this orbital distance maintain their mass over the lifetime of the system.

After the primordial atmosphere has been lost, the core of the planet is
exposed.  If the core of the planet is composed primarily of low-temperature
condensates (consistent with formation in the cold outer disk -- Pollack \etal
1996; Boss 1997), the outer layers of the planetary core may continue to
vaporize. To determine the original mass of a hot Earth formed from the core
of a photo-evaporated massive planet, models of the evolution of intensely
irradiated icy bodies must be developed; current models of volatile-dominated
Earth-mass planets (Valencia \etal 2006, 2007a, 2007b; Fortney \etal 2007;
Sotin \etal 2007; Seager \etal 2007) have not yet probed these processes.

Thus, photo-evaporation of close-in gaseous planets may remove their
atmospheres and leave behind solid cores.  This process is only effective for
planets within about 0.05 AU that are below the type 2 migration threshold of
$\sim 70 \mearth$.  For those cases, ``hot Neptunes" could become ``hot
Earths" on a $10^{8-10}$ year timescale.  This process can occur in
conjunction with the type 1 migration scenario discussed above in $\S$ 3.2.
Indeed, the most likely source of hot Neptunes is the outer disk (Gillon \etal
2007).  In some cases, a series of $\ga 10 \mearth$ planets may form in the
cold outer disk and then type 1 migrate inward as described above, into a
chain of hot Neptunes.  Depending on their orbits, the innermost planet or two
could be photo-evaporated over time into a very water-rich hot Earth.  A
diagnostic of photo-evaporation could therefore be a system with 1) a
water-rich hot Earth inside 0.05 AU and 2) additional, $\sim$ Neptune-mass
planets exterior to the hot Earth in near-resonant orbits.  However, it would
still be difficult to definitely assess the degree of photo-evaporation in such
a setting.

More definitive detections of the importance of photo-evaporation would require
statistics of a large number of hot Earths and hot Neptunes orbiting stars
with a range of ages.  A correlation between the number of hot Earths vs. hot
Neptunes and the stellar age would indicate that such planets were losing mass
in time, presumably via photo-evaporation.  Alternatively, exploring the mass
functions of close-in planets down to lower masses could reveal time-dependent
mass loss (as in Hubbard \etal 2007a).

\section{Origin of the Known Hot Earth Systems}

\subsection{Gliese 876}

Gliese 876 is a 0.32 $\msun$ star (M4 dwarf) less than 5 parsecs from the Sun
(Marcy \etal 1998).  Its known planetary system contains a $\sim 7.5 \mearth$ hot
Earth at 0.02 AU, as well as two additional, $\sim$Jupiter-mass planets in a 2:1
resonance on more distant orbits (Marcy \etal 1998, 2001; Rivera \etal 2005).
The separation between the hot Earth and the giant planets is significant: the
ratio of orbital periods between the hot Earth and inner giant planet is 16.6.
Both models and observations suggest that Jovian planets are rare around
low-mass stars (Laughlin \etal 2004; Ida \& Lin 2005; Endl \etal 2006; Butler
\etal 2006, Gould \etal 2006).  Thus, the existence of two such massive
planets around GJ 876 may indicate that its protoplanetary disk was
particularly massive (e.g., Wyatt \etal 2007; Lovis \& Mayor 2007).

Could the GJ 876 hot Earth at 0.02 AU have formed {\it in situ}?  If so, then there must
have been at least 7.5 $\mearth$ in solids interior to $\sim$0.05 AU, assuming
accretion was efficient.  In the minimum-mass solar nebula (MMSN) model,
assuming the surface density $\Sigma$ scales as $r^{-3/2}$, there is $\sim
0.75 \mearth$ inside 0.05 AU, assuming the disk to extend all the way into the
star (Weidenschilling 1977; Hayashi 1981; Raymond \etal 2007).  For a more
common disk profile of $\Sigma \propto r^{-1}$ (e.g., Andrews \& Williams
2007), and applying a MMSN prescription, there is only 0.06 $\mearth$ inside
0.05 AU.  Thus, if 7.5 $\mearth$ of material existed in the inner 0.05 AU of
GJ 876's disk, then that disk must have been 10-100 times more massive than
the solar nebula.  This value is rather large, but it is not outside the realm of possibility, given the large spread in observed disk masses (Andrews \& Williams 2005; Scholz \etal 2006).  However, such a massive disk would be an anomaly, and the fraction of disks that could form such a close-in planet is small (Raymond \etal 2007). In addition, given the $\sim$ linear relation between disk mass and stellar
mass (e.g., Scholz \etal 2006), such a massive disk is an additional three
times less likely.  In addition, given the large dynamical separation between
the hot Earth and the closest Jovian planet, there is no clear explanation for
the lack of additional hot Earths.  Thus, it is unlikely that GJ 876's hot
Earth formed {\it in situ}.

If the GJ 876 hot Earth formed at a distance and type 1 migrated inward, we would expect it to have companions of similar mass in near resonant orbits.  No such companions have been discovered to date, although planets of a few $\mearth$ would probably not be detectable (Rivera \etal 2005).  However, given the existence of the two Jovian planets, perhaps there was a limited window of time for type 1 migration into the inner disk: once the giant planets formed, they would pose a barrier for
smaller migrating bodies (Thommes 2005).  A mass of 7.5 $\mearth$ is
consistent with a single hot Earth migrating into the inner disk, then
stalling where the type 1 torques disappear (Masset \etal 2006b; Paardekooper
\& Mellema 2006).

Zhou \etal (2005) explain the origin of the GJ 876 hot Earth with a
combination of shepherding from giant planet migration and SR sweeping. In
Zhou \etal's model, the two Jovian planets formed on more distant orbits, and
were trapped in resonance during migration (e.g., Lee \& Peale 2002).  This
migration also induced the formation of planets inside strong resonances, by
the migration shepherding mechanism described in $\S$ 3.3.  The giants'
migration stalled close to their current orbits, but subsequent dissipation of
the disk induced SR sweeping, promoting further accretion and shepherding the
hot Earth farther away from the giant planets.  This two-step model does not
require as large a disk mass as {\it in situ} accretion, because some mass
from more distant regions is shepherded into the inner disk.  In addition, it
predicts a significant separation between the giant planets and the hot Earth,
caused by the SR sweeping after migration.  However, this model has some
uncertainties.  For example, the violent nature of star-forming environments
may cause episodic pulses in the evaporation of the disk (Adams \etal 2004)
and therefore in the location of SRs (Ward 1981).  In such non-monotonic SR sweeping, it is unclear if material can still be shepherded.

In Zhou \etal's model, it is likely that the hot Earth is relatively dry,
assuming its composition is determined by the formation zone of the innermost
giant planet, and that accretion followed roughly as in Mandell \etal (2007).
In the type 1 model, the hot Earth could be rocky or icy, also depending on
its formation zone.  If GJ 876 d were to transit its host star, then its bulk
composition (rocky vs. icy) could be determined (see $\S$ 2). If it were shown
to be icy in nature, that would support the type 1 migration scenario.
However, if it were rocky, it would lend support to Zhou \etal's model.

\subsection{Gliese 581}

Gliese 581 is a 0.31 $\msun$ M3 dwarf at a distance of 6.3 pc from the Sun
(Hawley \etal 1997).  Its planetary system contains three hot Earths/Neptunes
with orbits between 0.04 and 0.25 AU and minimum masses between 5 and 15
$\mearth$ (Bonfils \etal 2005; Udry \etal 2007).  The innermost planet is the
hot Neptune (M sin i = 15.7 $\mearth$).  No Jovian planets have been detected
in the system to date, ruling out the two shepherding mechanisms.

The most likely formation mechanism of the GJ 581 system is either {\it in
situ} formation or type 1 migration (S. Raymond \etal, in preparation).  The
orbital periods of the planets do not form an obvious pattern -- the period
ratios between planets b/c and c/d are 2.38 and 6.34, respectively.  The
semimajor axes of planets b/c and c/d are separated by 20.5 and 47 mutual Hill
radii, respectively, similar to values for the Solar System's terrestrial
planets.

For the GJ 581 planets to have formed {\it in situ} would require $\sim 40-50
\mearth$ inside 0.5 AU.  Indeed, the simulation from Fig.~\ref{fig:insitu} is
an attempt to reproduce the system via {\it in situ} accretion.  By the same
arguments as made above, this would require a disk that is, at least in its
inner regions, 17-50 times more massive than a minimum-mass disk.  Given that
the spacing of planets b, c and d is comparable to those of Venus, Earth and
Mars, {\it in situ} accretion remains a reasonable model for GJ 581.  In this
scenario, the innermost planet would have accreted first, and therefore may
have been able to capture a small amount of nebular gas to account for its
large mass (e.g., Pollack \etal 1996).

Formation at larger orbital distances followed by type 1 migration is the other viable mechanism for GJ 581.  The planets'
spacings are not next to obvious resonances, but b/c lie less than 10\% from
the 5:2 MMR and quite close to the 12:5.  Planets c/d are more distant, but of
course there exists the possibility of an additional, slightly lower-mass
planet between planets c and d.  If such a planet were discovered, it would
support the type 1 migration scenario.

Tidal effects are important in the GJ 581 system, given the planets' proximity
to the star.  Given that tides damp both semimajor axes and eccentricities, it
is likely that the GJ 581 planets b and c formed on more distant and more
eccentric orbits (Barnes \etal 2007).  Given the planets' already significant
eccentricities ($e \sim 0.2$ for each planet), it is not clear how the system
could form with such high eccentricities.  In addition, the fact that the
innermost planet is the most massive of the three suggests that
photo-evaporation has not occurred in this system.  Indeed, given the star's
low luminosity (1.3\% of solar), the threshold distance for photoevaporation
is likely to be at less than 0.01 AU.

Despite the uncertainties, the main difference between the two possible models
is simply the composition of hot Earths.  {\it In situ} formation predicts
relatively dry planets, while type 1 migration predicts icy planets with
$>10\%$ water by mass.  Thus, if transits were measured for any of the GJ 581
planets and a composition were determined, then it would be possible to distinguish between these two models.

\section{Summary and Discussion}

We anticipate that a large number of planetary systems containing close-in
terrestrial planets, referred to here as ``hot Earths'', will be discovered in
the coming years with radial velocity and transit measurements.  In some cases, both an accurate determination of the architecture of the inner planetary system and of the bulk composition of a hot Earth (rocky vs. icy; but see Adams \etal 2007) will be possible (see $\S$ 2).  The goal of
this paper is to determine whether the formation history of such systems can
be unraveled, given the relatively small amount of information available.  In
addition to four already-known mechanisms for hot Earth formation, we have
shown that tidal circularization of highly eccentric planets can move
terrestrial planets' orbits inward, but only by perhaps 0.1 AU, and only for
very close-in perihelion distances ($\la 0.05$ AU).  In addition, our simple
model suggests that photo-evaporation can remove a giant planet's atmosphere
and leave behind the core.  However, this is only possible for very close-in
orbits ($< 0.025-0.05$ AU) and relatively low-mass planets (``hot Neptunes"
with masses below $70 \mearth$), as suggested by Hubbard \etal (2007a).

Table 1 summarizes the observable consequences of these six models for hot
Earth formation.  There exist several clear differences between the models
that should be detectable in the near future.  Given a planetary system with a
transiting hot Earth, considerable RV measurements, and perhaps transit timing
analysis, Table 1 provides a simple way to determine the formation history of
hot Earth planetary system.  Note that in some cases, more than one of these
mechanisms can act in concert.  For example, the case of GJ 876 may be
explained in a two-step process, via shepherding during migration and then
during secular resonance sweeping (see $\S$ 4.1; Zhou \etal 2005).  In
addition, tides affect the orbits of all hot Earths to some degree, regardless
of their formation history.  However, certain mechanisms cannot act together:
planets massive enough to type 2 migrate cannot have their envelopes
photo-evaporated and become hot Earths (see $\S$ 3.6).

The formation mechanisms of the two known hot Earth systems are not entirely
clear (see $\S$ 4 above).  However, transit measurements of the hot Earth of
either of the known systems would make it far easier to discern between
models.  In particular, for the case of GJ 581, a transit measurement of
planet c or d would distinguish between {\it in situ} formation (rocky) and
type 1 migration (icy).  Clearly, more work is needed to better characterize
and quantify some of these models, and to examine the long-term survival of
hot Earths in different systems.  In addition, it is possible that additional
mechanisms exist for hot Earth formation that have not yet been considered.

Zhou \etal (2005) claimed that hot Earths should be numerous if giant planets
form via core-accretion (Mizuno 1980; Pollack \etal 1996; Lissauer \&
Stevenson 2007), but rare if they form via gravitational instability (Boss
1997; Mayer \etal 2002; Durisen \etal 2007).  Given the large number of
avenues for hot Earth formation, we disagree with Zhou \etal on this point.
Indeed, three of the candidate mechanisms for hot Earth formation -- {\it in
situ} accretion, type 1 migration, and tidal circularization -- do not
require a giant planet at all and so are unaffected.  Photo-evaporation of hot Neptunes may be affected because the two giant planet formation models predict different core masses: core-accretion predicts 5-20 $\mearth$ cores (e.g., Alibert \etal 2005) while the cores of giant planets formed via disk instability are likely to be smaller (Boss 1998, 2006).  For the other two mechanisms -- giant planet migration shepherding and secular resonance shepherding -- is there a reason that the outcome
should depend on the mode of giant planet formation?  The main difference between
the two models is the timing of giant planet formation: core-accretion
predicts that giant planets form late in the lifetime of the gaseous disk,
while gravitational instability forms planets very quickly.  Giant planet
migration starts immediately after, or even during, formation (Lufkin \etal
2004).  Thus, if giant planets form via core-accretion, they migrate through a
disk that has undergone at least 1 Myr of accretion, and contains both
$\sim$Moon-sized planetary embryos and km-sized planetesimals (plus $\sim$
99\% gas; e.g., Kokubo \& Ida 2000; Chambers 2004).  If, however, giant
planets form via gravitational instability, then they would migrate through a
disk containing predominantly smaller bodies such as planetesimals.  Fogg \&
Nelson (2005, 2007) showed that the prevalence of shepherding vs. scattering
during migration is relatively insensitive to the accretion history of the
inner disk.  In the secular resonance shepherding model, two giant planets
must be on interacting orbits by the late stages of the dispersal of the
gaseous disk; the planets' prior orbital histories are not relevant.  Thus, we
see no reason that the abundance or rarity of hot Earths should be affected by
the mechanism for giant planet formation.

One other interesting difference between the core accretion and gravitational
instability models is the expected location of giant planet formation.  In
core accretion, there are several reasons to expect giant planets to form just past the snow line: 1) the
density of solid building blocks increases by a factor of 2-4 or more (Hayashi 1981; Stevenson \& Lunine 1988; Lodders 2003), 2) accretion timescales are shorter than anywhere else beyond the snow line (Kokubo \& Ida 2002; Ida \& Lin 2004), and 3) the surface density jump at the snow line, if it is steep enough, can trap inward-migrating planetary cores and form a pileup (Masset \etal 2006b).  If these arguments hold, then core-accretion predicts that material interior to the giant planet is therefore
relatively dry.  However, gravitational instability forms planets in the more
distant reaches of protoplanetary disks, where the Toomre $Q$ value is lowest
(Boss 1997; Mayer \etal 2002).  Thus, icy material is included interior to the
giant planet.  In the giant planet migration shepherding model, hot Earths are
a mixture of material interior to the giant planet's starting orbit (Mandell
\etal 2007); they would be rocky for the core accretion model, and icy for the
instability model.  Thus, transit measurements of hot Earth in systems formed
by giant planet migration shepherding may provide a test to distinguish
between the two dominant giant planet formation models.

As observational uncertainties of planetary orbits and masses become smaller,
it will become possible to differentiate formation mechanisms based on these
observations. We have laid out the qualitative differences between six
different mechanisms that may form hot Earths (although some phenomena may
operate simultaneously). Determining how hot Earths form is an important step
toward understanding planet formation, identifying target stars for future
surveys, and searching for habitable planets.

\section{Acknowledgments}

We thank the reviewer, Jonathan Fortney, for constructive comments that improved the paper.
This work was motivated in part by a special session on ``hot Earths'' at the
210th AAS meeting in Honolulu May-June 2007, organized by Eric Gaidos and Nader Haghighipour.  We benefited from discussions with Bill Hubbard, Adam Burrows, Brian Jackson, Rick Greenberg, Dave Latham, Eric Agol, John Rayner, Jason Steffen, Eric Gaidos, and Ted von Hippel.  S.N.R. and A.M.M. were supported by appointments to the NASA Postdoctoral Program at the University of Colorado Astrobiology Center and NASA Goddard Space Flight Center, respectively, administered by Oak Ridge
Associated Universities through a contract with NASA.  R.B. acknowledges
supports from NASA's PG\&G and TPFFS programs.

\newpage
%\addcontentsline{toc}{section}{References}

\label{lastpage}

\end{document}